\documentclass[iop,twocolumnappendix]{emulateapj}
\usepackage{natbib}
\usepackage{aas_macros}
\usepackage{graphicx}
\usepackage[breaklinks,colorlinks,citecolor=blue]{hyperref}

\begin{document}

\title[Mapmaking with a messenger field]{Cosmic Microwave Background Mapmaking with a Messenger Field}

\author{Kevin M. Huffenberger,$^{1}$\thanks{khuffenberger@fsu.edu} Sigurd K. N{\ae}ss$^{2}$}

\affiliation{
  $^{1}$Department of Physics, Florida State University,
  $^{2}$Center for Computational Astrophysics, Flatiron Institute
}

\begin{abstract}
  We apply a messenger field method to solve the linear minimum-variance mapmaking equation in the context of Cosmic Microwave Background (CMB) observations.   In simulations, the method produces sky maps that converge significantly faster than those from a conjugate gradient descent algorithm with a diagonal preconditioner, even though the computational cost per iteration is similar. The messenger method recovers large scales in the map better than conjugate gradient descent, and yields a lower overall $\chi^2$.  In the single, pencil beam approximation, each iteration of the messenger mapmaking procedure produces an unbiased map, and the iterations become more optimal as they proceed.  A variant of the method can handle differential data or perform deconvolution mapmaking.  The messenger method requires no preconditioner, but a high-quality solution needs a cooling parameter to control the convergence.  We study the convergence properties of this new method, and discuss how the algorithm is feasible for the large data sets of current and future CMB experiments.
\end{abstract}
%
\section{Introduction}

 Cosmic Microwave Background (CMB) experiments measure the temperature and polarization of radiation from the $z\sim 1100$ surface of last scattering.  To reduce systematics, a CMB telescope scans the sky in constant motion.  As an important step in the analysis, we must assemble the scanning data into sky maps, but doing so is computationally expensive.  \citet{1992ASIC..359..391J} suggested a linear, minimum-variance solution to the mapmaking problem for COBE, and \citet{1997ApJ...480L..87T} showed that this solution preserves the information content of the data.  While there are other methods that also preserve information, this solution makes few assumptions and produces an unbiased map with minimized errors, which are appealing features.

As CMB experiments achieved higher resolution, the computational scaling and increased number of map pixels made brute force approaches to the minimum variance solution impractical, and researchers turned to iterative linear system solvers that could find the solution without explicitly performing an intractable matrix inversion \citep{1996ApJ...458L..53W,2001A&A...374..358D,2001A&A...372..346N, 2002MNRAS.330..497D, 2002PhRvD..65b2003S, 2005A&A...436.1159D}.  Preconditioned conjugate-gradient descent is the currently preferred method and these solvers have subsequently been optimized for massively-parallel high-performance computing environments \citep[e.g.][]{2010ApJS..187..212C}.

Preconditioners speed up the convergence to the solution.  These are approximate matrix inverses whose purpose is to lower the condition number of the matrix in the linear system.  With simple preconditioners, these methods typically converge in a few hundred iterations for CMB data sets.
With a more carefully design preconditioner, the number of iterations can be greatly reduced \citep{2014JCAP...08..045N,2014A&A...572A..39S}, but the preconditioner must be tuned carefully to fit the characteristics of the CMB survey (using e.g.\ a typical scanning pattern).  Such computations can be much more costly per iteration.

In this paper we introduce a new algorithm for map making, based on a messenger field method.  Messenger fields were introduced by \citet{2013A&A...549A.111E} to circumvent the inversion of a dense covariance matrix in the computation of the linear Wiener filter.  The messenger method is iterative, and  requires no preconditioner. Along with the map, it solves for an additional field (the ``messenger''), but only requires inversions of sparse matrices.  Beyond solving the linear system for the Wiener filter, the messenger field can be used to generate samples of the signal and signal covariance via Gibbs sampling.  Several subsequent works have applied the messenger method to problems in cosmology, including CMB, gravitational lensing, and large scale structure \citep{2015ApJ...808..152A,2015MNRAS.447.1204J,2016MNRAS.455.3169L,2016MNRAS.455.4452A}.  As a relatively new approach, it is still being refined to improve the convergence \citep{2017arXiv170208852K} and to treat wider varieties of covariance matrix structures \citep{2017arXiv170400865H}.

The approach to the mapmaking problem that we present here  uses messenger fields, but differs from the above works in two important respects.  First, we solve a different linear system in the mapmaking problem than in the Wiener filter.  Mapmaking makes no assumption about the signal's covariance.  Second, in all the above works, the data and the signal vectors have the same dimensionality, that is, the same number of pixels.  This is not true in mapmaking, since the time-ordered data for all detectors contains vastly more samples than the number of pixels in the temperature and polarization components of the resulting maps.
The time samples are tied to positions on the sky map by the pointing of the telescope's detectors.
Many time samples refer to the same map pixel, and the cross-linking of the scans is important to optimize the measurement of the sky.

The tests we report here indicate that our new messenger mapmaking algorithm can converge much faster than the preconditioned conjugate gradient descent method, at a similar computational cost per iteration.  Furthermore, the method is feasible for large data sets, an important feature for modern CMB experiments that utilize many thousands of detectors and multi-year campaigns.

This paper is organized so that  in  Section \ref{sec:methods} we introduce the messenger mapmaking method.  In Section \ref{sec:results}, we present the results of the method applied to simple simulations that allow a careful examination of the convergence properties.   In Section \ref{sec:discussion}, we discuss scaling up the method to real data sets.  In Section \ref{sec:conclusions}, we conclude with a brief description of a preliminary application to ACTPol data.  Three appendices provide derivations of the method for single pencil beam and for composite (e.g.\ differential) beam experiments, and also present an extension that can treat complications in the noise properties, but at increased computational cost.

\section{Methods}\label{sec:methods}
The standard model for a vector of time-ordered data is
\begin{equation}
  d = \mathbf{P} m_{\rm true} + n,
\end{equation}
where $\mathbf{P}$ is the pointing matrix that scans the sky (vector $m_{\rm true}$), and $n$ is zero-mean noise in the timestream.  The pointing matrix contains the weight that a particular time sample provides to a map pixel (in temperature and polarization).  The minimum-variance, unbiased estimate for the sky is given by the mapmaking equation,
\begin{equation}
  m = (\mathbf{P}^\dag\mathbf{N}^{-1} \mathbf{P})^{-1} \mathbf{P}^\dag \mathbf{N}^{-1} d, \label{eq:mapmaking}
\end{equation}
where $\mathbf{N}$ is the covariance matrix of the noise.
This solution minimizes
\begin{equation}
  \chi^2(m) = (d - \mathbf{P} m )^\dag \mathbf{N}^{-1} (d - \mathbf{P} m ),\label{eq:chi2}
\end{equation}
and therefore maximizes the Gaussian probability $\propto \exp(-\chi^2(m)/2)$.  It is also unbiased, $\langle m \rangle = m_{\rm true}$.  The noise covariance for data timestreams is approximately diagonal in the frequency domain, which makes the multiplication of the data by the inverse noise matrix tractable. However, the matrix inverse $(\mathbf{P}^\dag\mathbf{N}^{-1} \mathbf{P})^{-1}$ is not directly computable for such correlated noise and large numbers of pixels, and the problem requires a linear system solver.  Conjugate gradient descent methods, for example, move $(\mathbf{P}^\dag\mathbf{N}^{-1} \mathbf{P})$ to the left hand side of equation (\ref{eq:mapmaking}) and prescribe iterative updates to $m$ to find the solution.
 
\subsection{Messenger field mapmaking}

In an appendix, 
we show that iterating the following equations provides an alternative solution to the mapmaking equation:
\begin{eqnarray}
  t_i &=& \left( \bar \mathbf{N}^{-1} +  (\lambda\mathbf{T})^{-1} \right)^{-1} \left(\bar \mathbf{N}^{-1} d + (\lambda\mathbf{T})^{-1} \mathbf{P} m_i \right) \qquad \\ \nonumber
  m_{i+1} &=& (\mathbf{P}^\dag\mathbf{T}^{-1} \mathbf{P})^{-1} \mathbf{P}^\dag \mathbf{T}^{-1} t_i,
\end{eqnarray}
for $\bar \mathbf{N} = \mathbf{N} - \mathbf{T}$ and $\lambda = 1$.
This approach is inspired by the messenger field method of \citet{2013A&A...549A.111E}. The messenger field $t$ is a weighted average of the data and the map (projected into the timestream by the pointing matrix). 

We are free to choose any $\mathbf{T}$ that leaves $\bar \mathbf{N}$ positive definite. In the simplest implementations, we let $\mathbf{T} = \tau \mathbf{I}$.  In this specific case, $\mathbf{T}$ drops out of the second equation.  Since the identity matrix is diagonal in any basis, this choice makes all the matrix inversions trivial so long as time samples are each associated in $\mathbf{P}$ with one sky position only (the single, pencil beam approximation).  For $\mathbf{T} = \tau\mathbf{I}$, the messenger field $t$ is our statistical best estimate for a portion of the timestream that contains the signal and uniform, uncorrelated noise only, while the full timestream $d$ contains correlated noise as well.  For that case, the estimated map $m$ is a simple co-addition of $t$.  The value $\tau$ is the variance of the uniform noise in $t$, and so cannot be larger than the global minimum of noise in the data.  We will discuss $\lambda$, the ``cooling parameter,'' shortly.


Although the messenger field $t$ has this nice interpretation, it is large for real data sets, the same size as the time-ordered data.  It is preferable to precompute smaller data objects and discard the time-ordered data and similar objects from memory.  For the second equation we do not actually need $t$, just $\mathbf{P}^\dag\mathbf{T}^{-1} t$, and so we can combine the equations, rewriting the iteration using map-sized objects:
\begin{equation}
  m_{i+1} = m_d(\lambda) + \mathbf{F}(\lambda) m_i. \label{eq:mapsized}
\end{equation}
Here $m_d(\lambda)$ is a binned map of filtered data,
\begin{eqnarray}
  m_d(\lambda) &=& (\mathbf{P}^\dag \mathbf{T}^{-1} \mathbf{P})^{-1} \mathbf{P}^\dag\mathbf{T}^{-1} \left(\bar \mathbf{N}^{-1} + (\lambda\mathbf{T})^{-1}\right)^{-1} \bar \mathbf{N}^{-1} d \nonumber \\
  &=&  (\mathbf{P}^\dag \mathbf{T}^{-1} \mathbf{P})^{-1} \mathbf{P}^\dag (\lambda^{-1}  \bar \mathbf{N} + \mathbf{T})^{-1} d \label{eqn:md}
\end{eqnarray}
and $\mathbf{F}$ is a matrix filter applied to the current iteration,
\begin{eqnarray}
  \mathbf{F}(\lambda) &=& (\mathbf{P}^\dag\mathbf{T}^{-1} \mathbf{P})^{-1} \mathbf{P}^\dag \mathbf{T}^{-1} (\bar \mathbf{N}^{-1} + (\lambda\mathbf{T})^{-1})^{-1}(\lambda\mathbf{T})^{-1} \mathbf{P} \nonumber \\
  &=&  (\mathbf{P}^\dag\mathbf{T}^{-1} \mathbf{P})^{-1} \mathbf{P}^\dag \mathbf{T}^{-1} (\mathbf{I} + \lambda \mathbf{T} \bar \mathbf{N}^{-1})^{-1} \mathbf{P} \nonumber \\
  &=& \mathbf{I} - (\mathbf{P}^\dag\mathbf{T}^{-1} \mathbf{P})^{-1} \mathbf{P}^\dag(\lambda^{-1} \bar \mathbf{N} + \mathbf{T})^{-1} \mathbf{P}. \label{eqn:F}
\end{eqnarray}
The Woodbury formula leads to the last equation.  Together these yield another 
{compact form for the mapmaking algorithm, which is
\begin{equation}
  m_{i+1} = m_{i} +  (\mathbf{P}^\dag\mathbf{T}^{-1} \mathbf{P})^{-1} \mathbf{P}^\dag(\lambda^{-1} \bar \mathbf{N} + \mathbf{T})^{-1}(d- \mathbf{P}m_i), \label{eqn:mcompact}
\end{equation}
  although this form requires the full time-ordered data.
}

For practical map making, the main computational costs per iteration are the projection of the map into a timestream, the Fourier transform to the frequency domain for filtering by sparse matrices, the inverse Fourier transform back to the time domain, and the projection back to the map domain.  These costs per iteration are comparable to conjugate gradient descent with a diagonal preconditioner, but in our examples we find the messenger field method can converge to a higher-quality solution in many fewer iterations.

For general values of the cooling parameter $\lambda > 1$, the iterations instead minimize
\begin{equation}
    \chi^2(m,\lambda) = (d - \mathbf{P} m )^\dag \mathbf{N}'(\lambda)^{-1} (d - \mathbf{P} m )
\end{equation}
where $\mathbf{N}'(\lambda) = \mathbf{N} + (\lambda-1) \mathbf{T}$.  This minimization leads to the $\lambda$-dependent, unbiased map solution
\begin{equation}
  m(\lambda) = (\mathbf{P}^\dag \mathbf{N}'(\lambda)^{-1} \mathbf{P})^{-1} \mathbf{P}^\dag \mathbf{N}'(\lambda)^{-1} d, \label{eqn:mlambda}
\end{equation}
which is optimal only for $\lambda = 1$.

 For very large $\lambda$, $\mathbf{F}=0$ and the next iteration gets no contribution from the current one.  Then the map is an unbiased weighted map of the data: $m(\mbox{large } \lambda) = m_d = (\mathbf{P}^\dag\mathbf{T}^{-1} \mathbf{P})^{-1} \mathbf{P}^\dag \mathbf{T}^{-1} d$.  For $\mathbf{T} = \tau \mathbf{I}$, this is a co-addition of $d$.  Regardless of $\lambda$, if an iteration is unbiased ($\langle m_i \rangle = m_{\rm true}$), the messenger field has mean $\langle t_i \rangle = \mathbf{P}m_{\rm true}$, and the next iteration will be unbiased as well.

The residual in a map compared to the solution at a particular $\lambda$ is 
\begin{equation}
  \epsilon_{i+1}(\lambda) = m_{i+1}(\lambda) - m(\lambda) = \mathbf{F}(\lambda) \epsilon_i(\lambda).
\end{equation}
Thus the convergence of the remaining residual to zero depends on the eigenvalues of matrix $\mathbf{F}$.  If an eigenvalue is much less than unity, the corresponding eigenmode will converge fast. If it is close to unity, the mode will converge slowly.
Looking at the structure of $\mathbf{F}$, we may expect that this transition point will occur  (for frequency domain noise) around $N(f) \approx \lambda \tau$, so that modes less noisy than that will rapidly converge to their proper values for $m(\lambda)$.  


For modes that correspond to frequencies that are much more noisy than $\lambda \tau$, the solution $m(\lambda)$ and our desired $m(\lambda = 1)$ are the same.  Following \citet{2013A&A...549A.111E}, we thus establish a ``cooling schedule'' for $\lambda$ that takes it from large values to $\lambda = 1$ as we iterate.  We must lower the value of $\lambda$ so that less-and-less noisy modes converge to their final values, but not too quickly, or noisy modes will get stuck with eigenvalues close to unity before converging. A proper cooling schedule will permit noisy modes to converge quickly at large $\lambda$, and then let less noisy modes converge to their proper values as $\lambda \rightarrow 1$, speeding the convergence to the ultimate solution.

\subsection{Composite pointing matrix}
The above discussion of messenger mapmaking supposes that the matrix $\mathbf{P}^\dag\mathbf{T}^{-1}\mathbf{P}$ is simple to invert for suitable $\mathbf{T}$.  This is true in the single pencil beam approximation, which has only one pixel entry (or polarized pixel block) in the pointing matrix per time sample.  However, it is not true for differential experiments like COBE-DMR or WMAP, which sample and difference widely-separated directions on the sky.  In the differential case the pointing matrix has a $+1$ and a $-1$ pixel entry per time sample in the pencil beam approximation, corresponding to the pointing of the two sides of the differencing assembly \citep{1996ApJ...458L..53W,2003ApJS..148...63H}.  The pointing matrix has even more entries when attempting deconvolution mapmaking, where a portion of the beam is placed in the pointing matrix and deconvolved in a regularized way \citep{2004PhRvD..70l3007A,2009ApJS..181..533A,2010ApJS..187..212C}.

The case with two components has the form
\begin{equation}
  \mathbf{P} = \mathbf{P}_A + \mathbf{P}_B,
\end{equation}
but we could generalize this to an arbitrary number of components.  In the differential case  the $A$-component could have $+1$ entries and the $B$-component could have $-1$ entries.  We also split the messenger covariance as $\mathbf{T} = \mathbf{T}_A + \mathbf{T}_B$.  This form allows us to make progress because matrices like $\mathbf{P}_A^\dag\mathbf{T}_A^{-1}\mathbf{P}_A$, and sums of such matrices, can be made simple to invert.

In an appendix, we show that the messenger mapmaking solution in this case still has the form $m_{i+1} =  m_d(\lambda) + \mathbf{F}(\lambda) m_i$, with slightly modified terms (compare equations \ref{eqn:md} and \ref{eqn:F}):
\begin{eqnarray}
   m_d(\lambda) &=& (\mathbf{P}_A^\dag \mathbf{T}_A^{-1}\mathbf{P}_A + \mathbf{P}_B^\dag \mathbf{T}_B^{-1}\mathbf{P}_B )^{-1}\\ \nonumber & & \hspace{0.4\columnwidth} \mathbf{P}^\dag (\lambda^{-1} \bar \mathbf{N} +  \mathbf{T} )^{-1} d \\ \nonumber
  \mathbf{F}(\lambda) &=& \mathbf{I} -  (\mathbf{P}_A^\dag \mathbf{T}_A^{-1}\mathbf{P}_A + \mathbf{P}_B^\dag \mathbf{T}_B^{-1}\mathbf{P}_B )^{-1} \\ \nonumber & & \hspace{0.4\columnwidth} \mathbf{P}^\dag ( \lambda^{-1}\bar \mathbf{N} +  \mathbf{T} )^{-1} \mathbf{P}.
\end{eqnarray}
The matrix $(\mathbf{P}_A^\dag \mathbf{T}_A^{-1}\mathbf{P}_A + \mathbf{P}_B^\dag \mathbf{T}_B^{-1}\mathbf{P}_B)$ is a weighted sum of weight maps, and as before can be made (block) diagonal for simple $\mathbf{T}$ matrices.
Another compact form for the mapmaking solution is:
\begin{eqnarray} \label{eqn:mdiffcompact}
   m_{i+1} &=&  m_{i} + (\mathbf{P}_A^\dag \mathbf{T}_A^{-1}\mathbf{P}_A + \mathbf{P}_B^\dag \mathbf{T}_B^{-1}\mathbf{P}_B )^{-1} \\ \nonumber & & \hspace{0.3\columnwidth} \mathbf{P}^\dag ( \lambda^{-1}\bar \mathbf{N} +  \mathbf{T} )^{-1} (d - \mathbf{P}m_i),
\end{eqnarray}
which recalls equation (\ref{eqn:mcompact}).  

There are some important differences to the single beam case.  Here $\mathbf{F} \neq 0$ for large $\lambda$, so even in that case, the initial condition matters.  Unlike before, large $\lambda$ does not automatically ensure an unbiased map.  In some cases, for example when $\mathbf{T}_A = \mathbf{T}_B = \mathbf{T}/2$ and all are diagonal, then $(\mathbf{P}_A^\dag \mathbf{T}_A^{-1}\mathbf{P}_A + \mathbf{P}_B^\dag \mathbf{T}_B^{-1}\mathbf{P}_B )$ is proportional to the diagonal part of $ \mathbf{P}^\dag \mathbf{T}^{-1}\mathbf{P}$.  Then, in the limit of large $\lambda$, equation (\ref{eqn:mdiffcompact}) expresses a weighted Jacobi method that will converge toward the unbiased binned map solution $(\mathbf{P}^\dag \mathbf{T}^{-1}\mathbf{P})^{-1}\mathbf{P}^\dag \mathbf{T}^{-1} d$.  At other $\lambda$ the map will converge toward an unbiased but non-optimal $m(\lambda)$ (see equation \ref{eqn:mlambda}).  If the method achieves an unbiased map at any $\lambda$, or is initialized with an unbiased map, subsequent iterations will stay unbiased.

\begin{figure*}
  \includegraphics[width=\textwidth]{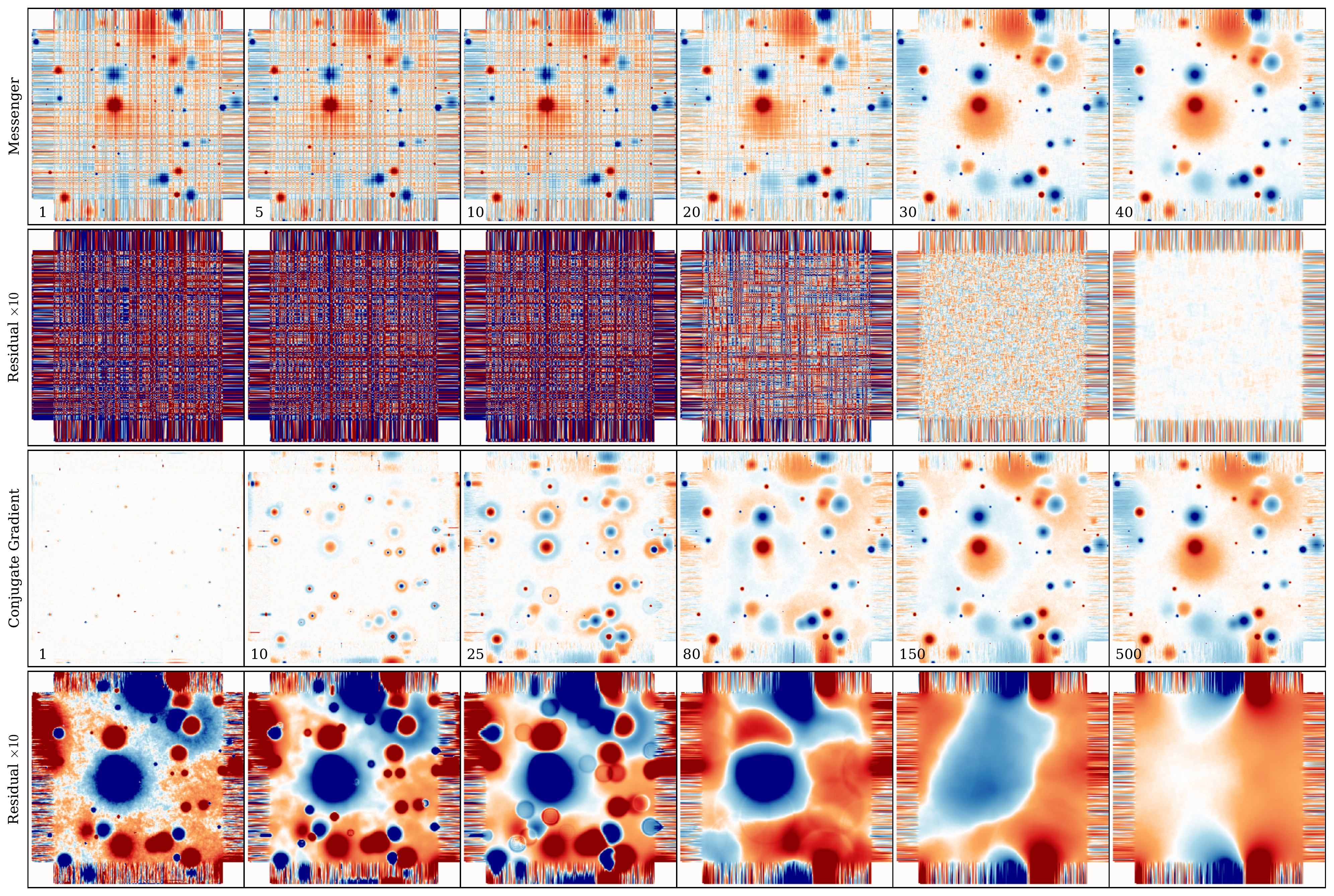}
  \caption{Convergence of messenger and conjugate gradient maps in the single beam case.  The top row shows the messenger mapmaker up to 40 iterations, using a cooling schedule with 8 cooling levels of 5 iterations each (denoted elsewhere $8\times5$).  The iteration number is noted on each panel.  The color scale has a range of $\pm 1000$ $\mu$K.  The second row shows the residual difference to a well-converged solution to the map (using the $16 \times 10$ cooling schedule), inflated by a factor of ten.   The third row show a conjugate gradient mapmaker with a diagonal preconditioner up to 500 iterations.  The bottom row shows the conjugate gradient residual differences to the well-converged map (inflated by factor ten).  The conjugate gradient map retains spurious large-scale features even at 500 iterations.  
  These images show the intensity component, but the polarization components show similar features.  }\label{fig:map-comparison}
\end{figure*}

\begin{figure}
  \includegraphics[width=\columnwidth]{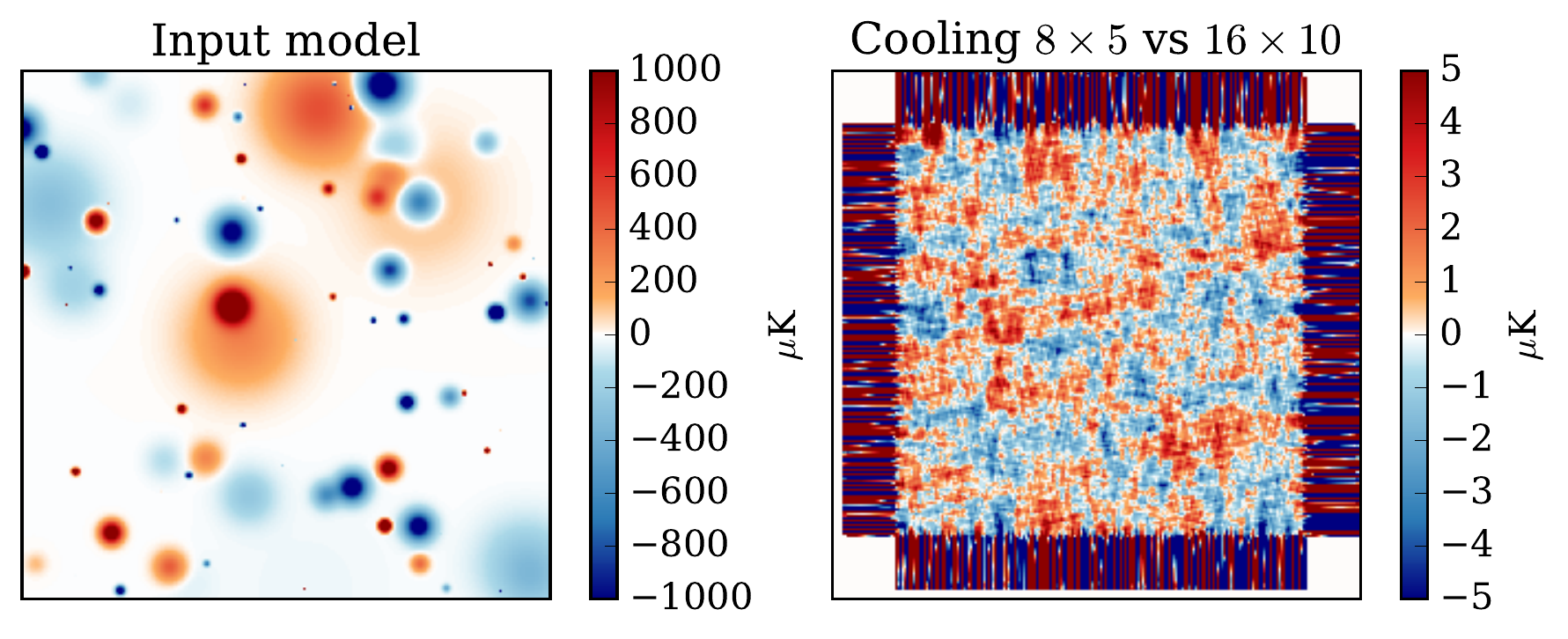}
  \caption{\textit{Left:} intensity component of our polarized model sky. \textit{Right}: difference between maps produced by messenger mapmaking with two cooling schemes.  Here we show the map with 8 cooling levels of 5 iterations each minus the more-converged map with 16 cooling levels of 10 iterations each.}\label{fig:model}
\end{figure}

\subsection{Simulations}

To validate and study the messenger mapmaking method,  we prepared mock time-ordered data based on cross-linked raster scans of model instruments over a simulated sky.    The scanned area is square and two degrees across.  The statistics of the sky signal are not important for the mapmaking solution, which depends only on the noise covariance and the pointing.  For the signal in the test case, we used a collection of compact and extended axisymmetric  polarized and unpolarized sources.  This lets us observe how different scales converge.  The source amplitudes range from a few hundred microkelvin to a few  millikelvin and their signals also vary within pixels.  All the map-making algorithms we tested here were subject to the same sub-pixel signal bias, so it does not figure into the relative comparisons. 

The single beam instrument model uses an orthogonal pair of raster scanning patterns, and has nine polarization sensitive detectors, at various orientations, that are offset from the telescope boresight by up to 1.8 arcmin.  The noise for each detector is uncorrelated and has a power spectrum with power-law and white noise components,
\begin{equation}
N(f) = N_0 \left(1+(f/f_{\rm knee})^{-2} \right),
\end{equation}
where $N_0 = (20$~$\mu$K)$^2$ and $f_{\rm knee} = 66.7$~Hz.  Each raster scan lasts for $1000$~seconds and covers the bulk of the scanning area.  The scans are unphysical: the boresight moves in one direction only without turn-arounds, and jumps back to the start with a frequency 0.46 Hz (every 2.17 seconds), while stepping ahead monotonically in the cross-scan direction at the end of each scan.

We made a second set of simulations to test the differential mapmaking case.  In this test we simplified the differential model instrument to have a single temperature-only detector (with two beams).  We offset the $A$-side and the $B$-side beams from the boresight in opposite directions, each by 9 arcminutes.  We scanned over the same model as before, fixing the angle of the boresight so that the separation between the beams is perpendicular to the scanning direction.  As a simplifying assumption, we discarded subpixel structure.  At the mapmaking resolution we used for the single-beam model, we find that the differential scenario is not well-crosslinked for the same pair of orthogonal scans.  Therefore we rotated the raster pattern to nine different orientations, shortening the angular length of each scan so that they  still fit within the two degree square of the model.  We adjusted the noise level of the detectors so that we spend the same total effort for both the differential and single-beam cases, although the single-beam scan pattern distributes it more evenly.

\section{Results} \label{sec:results}

\begin{figure}
  \includegraphics[width=\columnwidth]{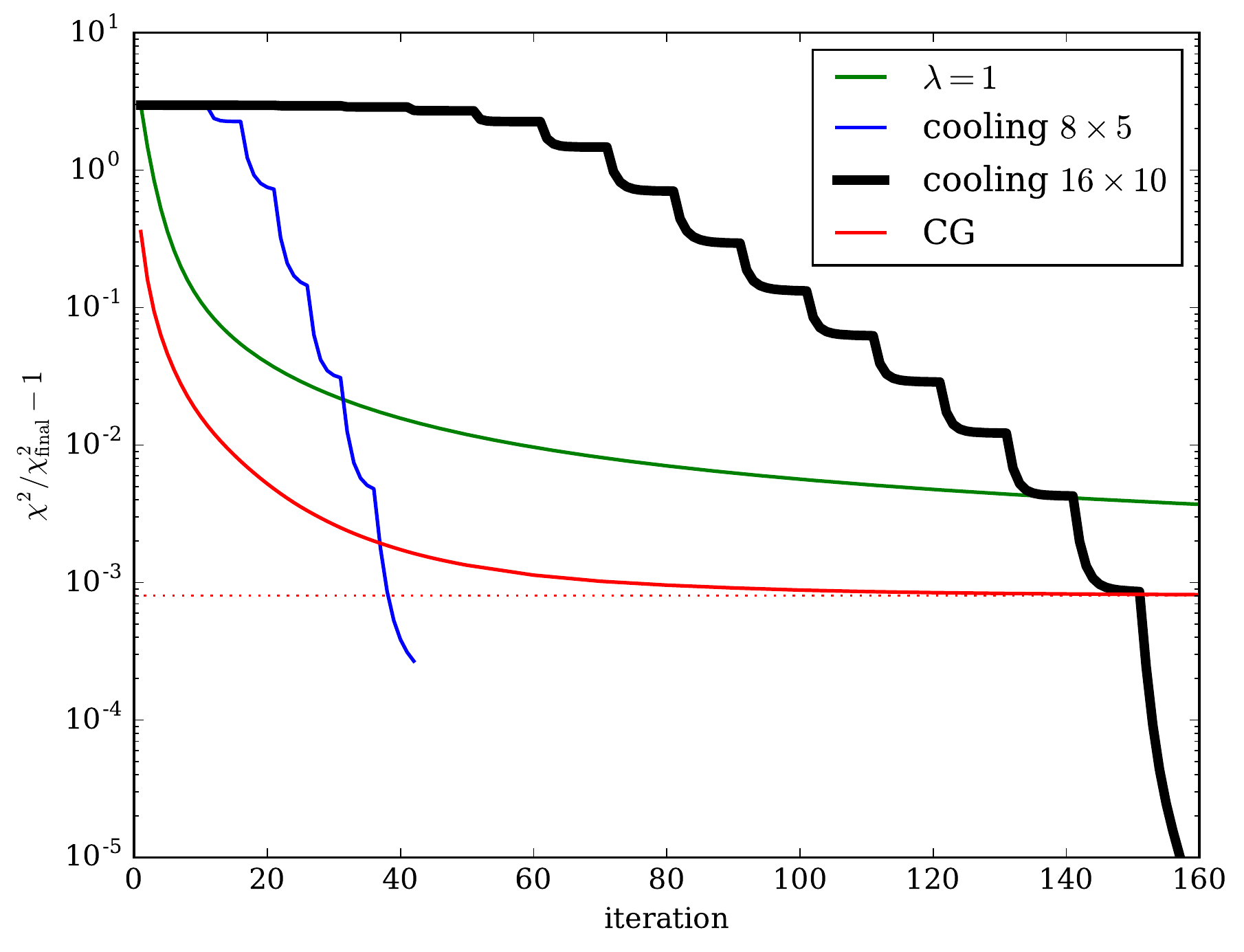}
  \caption{With an appropriate cooling schedule, the messenger method can achieve lower $\chi^2$ than the conjugate gradient algorithm (CG), in many fewer iterations.
    The plot shows the minimized quantity $\chi^2(m)$ (equation \ref{eq:chi2}) for several mapmaking schemes.  Lower values mean better maps.
    The green line shows the result of a messenger mapmaker with no cooling scheme, holding the cooling parameter constant at $\lambda=1$.  Two cooling schedules for $\lambda$ achieve better final outcomes.  Here we show 8 cooling levels of 5 iterations each ($8 \times 5$)  and 16 levels of 10 iterations each ($16 \times 10$).
    The red line shows the progress of a conjugate gradient method, and the dotted red line shows the $\chi^2$ after 500 iterations.   We compare all curves to the final $\chi^2$ from the $16 \times 10$ cooling scheme.
  }\label{fig:chi2_vs_iteration}
\end{figure}

\begin{figure*}
\includegraphics[width=\textwidth]{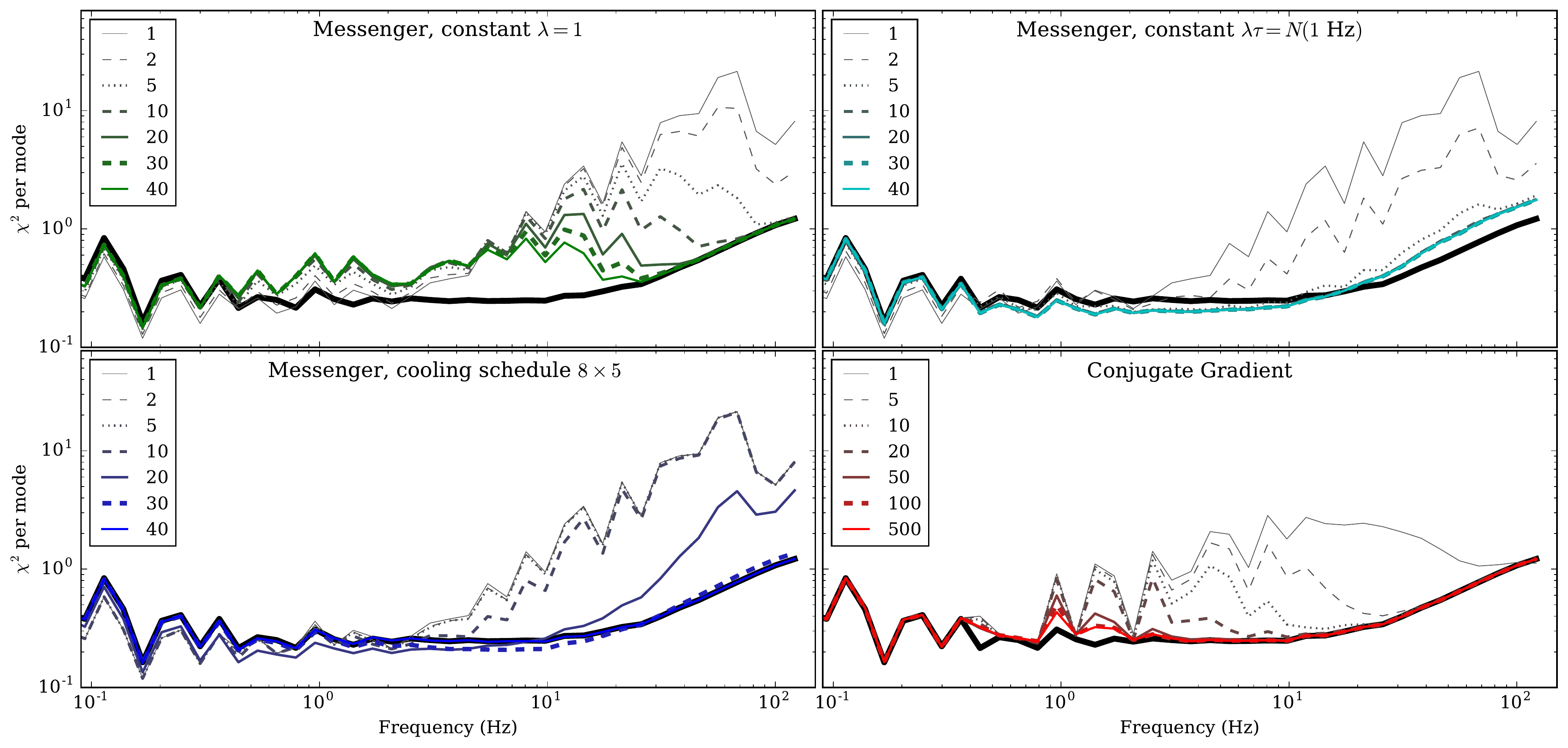}
\caption{Evolution of $\chi^2$ per frequency mode with iterations (considering one detector during one raster scan).  The color and line style indicate the iteration count, and the thick black line shows the $\chi^2$ values for the well-converged $16 \times 10$ cooling schedule for the $\lambda$ parameter.  The black line can be viewed as the target for convergence.    A logarithmic binning averages over modes for plotting.  The top row displays the convergence of the messenger method with a fixed $\lambda$ parameter, and shows how different frequencies converge in a case with no cooling ($\lambda = 1$) and a case with a higher $\lambda$. The bottom row contrasts a cooled messenger method to a conjugate gradient mapmaker with a diagonal preconditioner.  In the messenger method with the $8 \times 5$ cooling schedule, the high frequencies converge last as $\lambda$ drops.  The messenger method converges faster and more completely than the conjugate gradient mapmaker, which has not converged at $\sim 1$~Hz after 500 iterations.}\label{fig:chi2_per_mode}
\end{figure*}

\subsection{Single beam}
In Figure \ref{fig:map-comparison}, we compare the results of the messenger mapmaking procedure to a conjugate gradient mapmaker on the same time-ordered data.  The two algorithms converge to nearly the same endpoint, but in strikingly different ways.  The messenger method starts after one iteration with an unbiased binned map full of scanning stripes.  As the iterations proceed, the maps remain unbiased and become more optimal.  The stripes fade away, in this case after a few tens of iterations, and the map in the cross-linked region resembles the input map (Figure \ref{fig:model}).  This map used a cooling schedule that we discuss below.

The conjugate gradient maps converge much more slowly on large scales.  Initialized to zero, these maps start highly filtered, showing only the minima and maxima positions of the final map.  As the iterations proceed, the maps gradually become less biased.  A pulse of signal propagates from each pixel (most obviously from the bright extrema), filling out the rest of the map.  Even at 500 iterations, however, the map has large scale distortions compared to the messenger map and the input map.  These tend to emanate from the uncrosslinked regions.  There are also other narrow glitches at the map edges that are not present in the messenger map at 40 iterations. 

The mapmaking procedure is supposed to minimize $\chi^2(m)$ (equation \ref{eq:chi2}).  In Figure \ref{fig:chi2_vs_iteration}, we show that $\chi^2$ decreases monotonically for iterations of both the messenger and conjugate gradient map makers.  We evaluate $\chi^2$ in the Fourier domain as a sum of contributions from every Fourier mode for all detector timestreams for all scans.  The conjugate gradient maps have a $\chi^2$ that drops rapidly, but reaches a plateau.

The convergence for the messenger method depends strongly on the cooling schedule. With no cooling schedule, at constant $\lambda = 1$, the $\chi^2$ falls quickly, but the noisy modes are slow to converge, and it would take hundreds or thousands of iterations to achieve a quality solution.

We implement cooling schedules where $\lambda$ is set very high for one iteration, which starts us with an unbiased binned map.  Therefore subsequent iterations are also unbiased.  Then $\lambda$ is set so that the levels of $\lambda \tau$ are logarithmically distributed between the maximum and minimum noise power, and we repeat each level for a fixed number of iterations.  This naturally concludes with $\lambda = 1$ for several iterations to finish the solution.  Thus the schedule with eight levels of cooling, for five iterations each, we denote as $8 \times 5$.  At a particular value of $\lambda$, the messenger maps will reach a plateau in $\chi^2$ after a few iterations as the low noise modes rapidly converge.  If there are noisy modes left to converge, they will  proceed slower, and $\chi^2$ falls more and more slowly.  When we lower $\lambda$, the $\chi^2$ rapidly falls toward the new equilibrium.  Cooling schemes can lead to better overall $\chi^2$ values than the conjugate gradient solution, in fewer iterations, and lead to better maps, particularly on the large scales that converge at high $\lambda$.

For this simulation, we argue that the $16 \times 10$ cooling scheme is well-converged, based on comparisons among several cooling schemes and the fact that the plateaus in $\chi^2$ are very flat, indicating that noisy modes were well converged at previous $\lambda$ levels.  In the $8 \times 5$ cooling scheme, the plateaus are not flat, so more convergence could be achieved at each $\lambda$ level.  Even so, the $8 \times 5$ cooling scheme after 40 or so iterations achieves a lower final $\chi^2$ than the conjugate gradient algorithm after 500 iterations.

\begin{figure}
\includegraphics[width=\columnwidth]{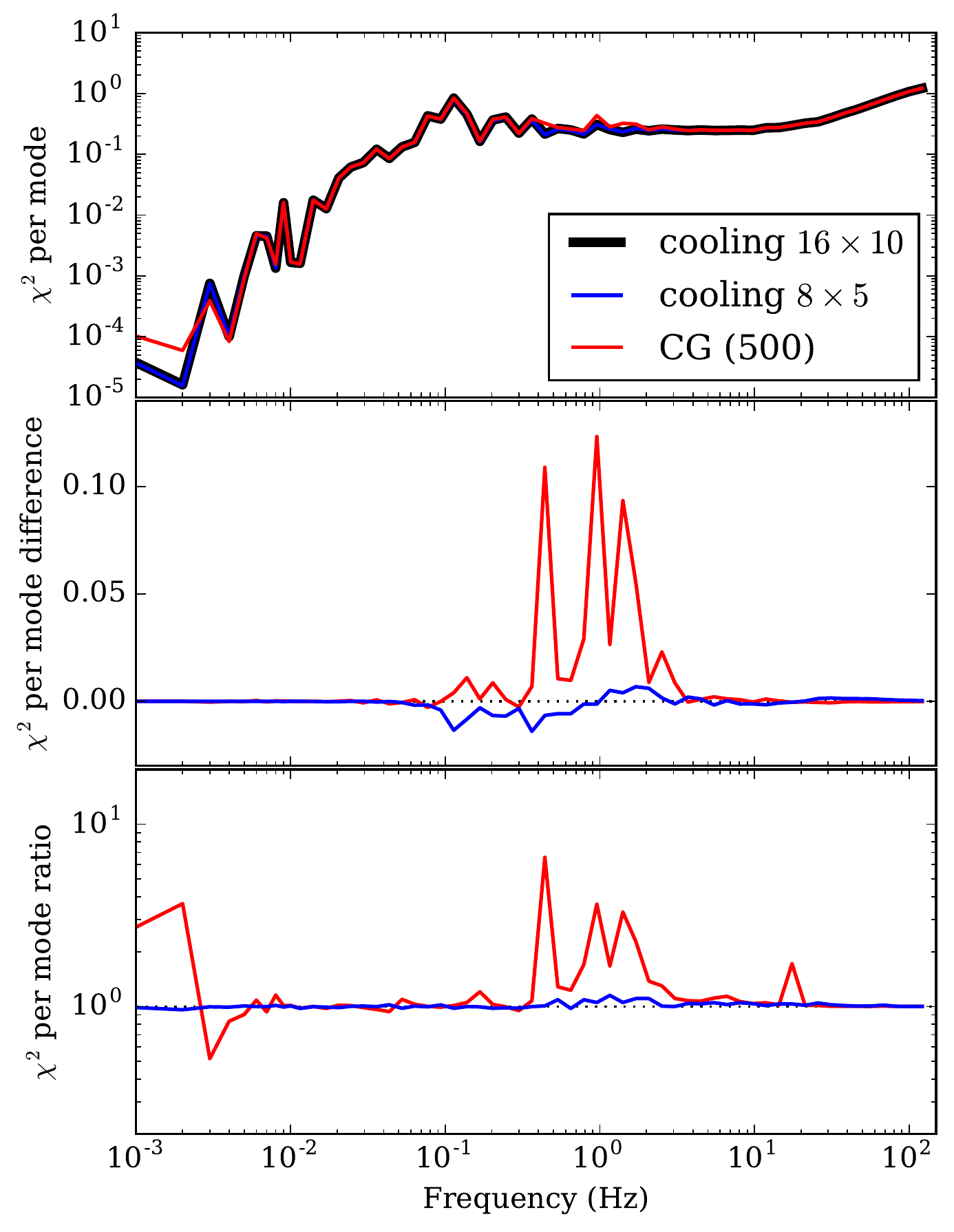}
\caption{Contrasting $\chi^2$  per mode for messenger and conjugate gradient mapmaking, for the timestream of a single detector during one raster scan of our simulation.  A logarithmic binning averages together nearby frequencies above 0.01 Hz.  At $<0.01$ Hz, we show individual modes.  The top plot compares the  $\chi^2$  per mode for the end point of the messenger map making with two cooling schemes, and for the conjugate gradient map maker after 500 iterations.
  The middle plot shows the difference to the most-converged $16 \times 10$ cooling schedule.  The bottom plot shows the ratio to the  $16 \times 10$ cooling schedule.  We divide before averaging into bins.}\label{fig:chi2_per_mode_ratio}
\end{figure}

Although $\chi^2$ is a good diagnostic, it does not tell us what parts of the solution are converging.  It is instructive to examine where the contributions to $\chi^2$ are coming from.  In Figures \ref{fig:chi2_per_mode} and \ref{fig:chi2_per_mode_ratio}, we examine $\chi^2$ per Fourier mode for a single detector during a single scan.  In the ideal limit, where the scan is completely in the cross-linked area and there is so much data that $m \approx m_{\rm true}$, the $\chi^2$ per mode for a segment of time-ordered data should be $\chi^2$-distributed with two degrees of freedom (for the real and imaginary portions of the Fourier coefficient) but rescaled to have unit mean.  We are not in that limit here, and the distribution of $\chi^2$ values per mode is a complicated interplay of the noise and the cross-linking of the scanning pattern.  It varies as a function of frequency, and has a substantial contribution from subpixel bias.

There is also no simple relationship between the frequencies in the time ordered data and the eigenmodes of $\mathbf{F}$ that converge in the map, except for a rough correspondence that high frequencies in the timestream tend to produce small scale features in the map, and low frequencies tend to produce large scale features (but not exclusively).

In Figure \ref{fig:chi2_per_mode}, for messenger iterations with constant $\lambda = 1$ (no cooling), we can see that the high frequencies find their final $\chi^2$ values first.  Lower and lower frequencies follow, but progress slows down as we are digging into noisier and noisier modes.  These plots are binned in frequency to average over nearby $\chi^2$ values, giving a rough idea of the mean of the distribution of $\chi^2$ as a function of frequency.  They are compared to the well-converged $16 \times 10$ cooling scheme.

In another panel we show iterations at a different fixed value for $\lambda$.  Here $\lambda \tau = N(1$~Hz$)$, and it converges toward the solution for that $\mathbf{N}'(\lambda)$.  For that solution, the $\chi^2$ per mode approaches its final value for very low frequencies.  At  $\sim 1$~Hz the $\chi^2$ per mode is lower than in the fully converged solution, while it is higher at high frequencies.  Essentially, this solution over-fits the $\sim 1$~Hz component of the noise and allows the high frequency part of the map to be too noisy compared to the ultimate ($\lambda = 1$) solution.

We also compare the $\chi^2$ per mode for a cooling schedule and a conjugate gradient mapmaker.
In the $8 \times 5$ cooling schedule solution, the early, higher-$\lambda$ iterations also have low $\chi^2$ at low frequency, but as $\lambda$ approaches unity the iterations force the $\chi^2$ of the low frequency modes up to their final values.  This balances the penalty in $\chi^2$ between low and high frequencies.
Compared to the others, the conjugate gradient solver converges very quickly at high frequencies, but even after 500 iterations, it does not converge at frequencies around 1~Hz.

In Figure \ref{fig:chi2_per_mode_ratio} we compare the final $\chi^2$ per mode for the messenger and conjugate gradient algorithms, extending to even lower frequencies.
We can see more clearly that the conjugate gradient solution's average $\chi^2$ per mode is above those for the pair of messenger solutions, and that even the relatively quick $8 \times 5$ cooling schedule provides a nearly converged map by comparison.  We note that the worst frequencies for the conjugate gradient method are near to the telescope scanning frequency (0.46 Hz), and also at the very lowest frequencies in the data.

We repeated this exercise with another simulation where all nine detector pointings coincided with the telescope boresight, and sampled pixel centers exactly, eliminating the effect of subpixel bias.  The messenger maps here are indeed unbiased, with no apparent residuals from the sky signal.  Without subpixel bias, the overall $\chi^2$ values are substantially lower.  The comparisons to the conjugate gradient maps lead to the same conclusions as before: conjugate gradient maps show large scale features that are not present in the messenger maps with cooling.  Like before, the $16 \times 10$ cooling scheme has the lowest $\chi^2$.  Unlike before, the $8 \times 5$ cooling scheme has a slightly higher $\chi^2$ than the conjugate gradient maps (at 500 iterations), but if the messenger algorithm is allowed to continue at $\lambda = 1$, it achieves a lower $\chi^2$ after 46 total iterations.

\subsection{Differential beams}
\begin{figure*}
  \includegraphics[width=\textwidth]{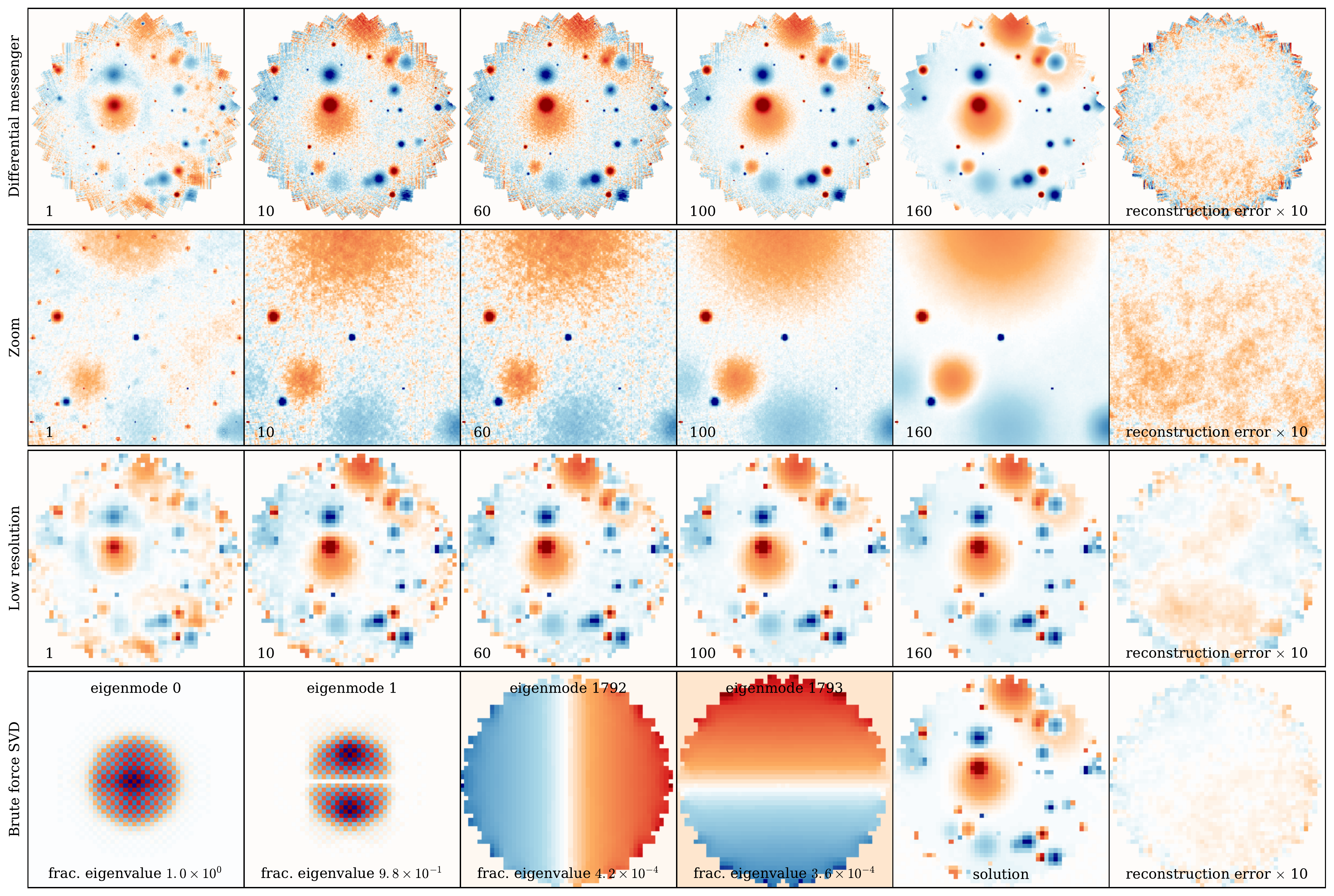}
  \caption{Messenger mapmaking algorithm for a differential instrument.  The top row shows iterations from a $16 \times 10$ cooling schedule. The color scale is $\pm 1000$~$\mu$K.  The last column shows the reconstruction error (compared to the input model) inflated by a factor of ten. (The mean is subtracted because it is not constrained in a differential experiment.)   The second row zooms in on a bright source that is below and left of the image center.  The first messenger iteration shows a ring of shadow beam images around the source, a bias that results from the differential beam  and scanning pattern.  There are two shadows for each of the nine scanning directions.  The shadows fade after a few iterations, and then the map gradually becomes more optimal as the cooling schedule proceeds.  The third row shows the same situation at lower resolution. The bottom row shows a brute force solution by singular value decomposition, which the low resolution permits.  We show the two most and two least constrained eigenmodes, the solution itself, and the reconstruction error to the input model. The $\chi^2$ of the low-resolution messenger solution is a fraction $10^{-3}$ higher than the brute force solution, and contains some additional large-scale noise fluctuations.} \label{fig:diffmess}
\end{figure*}

To verify our solution for a differential experiment, we implemented equation (\ref{eqn:mdiffcompact}) using $\mathbf{T}_A = \mathbf{T}_B = \mathbf{T}/2$, and applied it to our differential beam simulation.  We solve for the map and examine the residuals compared to input model.  (Recall that we discarded subpixel structure for this case.)  For the solution in Fig.~\ref{fig:diffmess}, we take one step at very high $\lambda$, then take $\lambda \rightarrow 1$ with a $16 \times 10$ cooling schedule.

In these examples, the effect of the differential beams is visible as a bias in the first iteration of the map.  Bright objects are surrounded by a ring of shadow images (of opposite sign).  These show the positions of the other beam when one beam has scanned over the bright object.  This example has nine scans of two beams, so there are eighteen shadows from each bright object.  All structures in the map are surrounded by similar features, which for extended objects blend together into doughnut-like shapes.

There are two distinct phases in the iterative solution.
First, in the early iterations, the shadow features fade away, and this debiasing dominates the reduction of $\chi^2$. In this example, the shadows have mostly disappeared on small scales after a dozen iterations, and the map becomes nearly unbiased.  Second, the subsequent iterations mostly serve to make the reconstructed map more optimal.  After 160 iterations, our solution in Fig.~\ref{fig:diffmess} has eliminated much of the noise.  The reconstruction errors that remain on medium to large scales appear to be noise, and change when varying the noise realization.

A test that narrows the separation between the differential beam pair from 18 arcminutes to 4.8 arcminutes provides a solution that is unbiased on small scales, but has a large scale bias that persists after a $16 \times 10$ cooling schedule, and does not appear to depend on the particular noise realization.  Running for more iterations at high $\lambda$ before starting the cooling schedule helps to reduce that bias, and so we conclude that the bias at least in part comes from a schedule that cools prematurely, before the map became unbiased on large scales, and erroneously freezes in the beam shadows of the large scale features.

Another possibility we need to consider is missing modes.  Depending on the cross-linking, differential experiments may struggle to represent certain modes on the sky.  The mean of the map is always unconstrained, for example, and other modes could be unconstrained also, particularly on scales larger than the beam separation.  Equivalently, this means the matrix $(\mathbf{P}^\dag\mathbf{N}^{-1} \mathbf{P})$ has less than full rank and so is not invertible.  To probe particular modes of the differential solution more carefully, we wanted to examine a brute force solution, but this is only practical at very low resolution.  We downsampled the model from the original map resolution of $600\times600$ pixels to $50\times50$ pixels.  We then generated time-ordered data from that map (with no subpixel structure) using the same scan pattern and noise realization.   Fig.~\ref{fig:diffmess} shows the messenger solution at low resolution with the same cooling scheme.  The messenger solution displays reconstruction errors that are similar to the high resolution solution.

At this lower resolution, we computed the brute force solution by tabulating the $2500^2$ entries of $(\mathbf{P}^\dag\mathbf{N}^{-1} \mathbf{P})$ explicitly.  From that matrix we computed eigenvectors, eigenvalues, and the Penrose pseudoinverse via singular value decomposition.  We find that 1794 eigenmodes have eigenvalues much larger than the others.  Large eigenvalues indicate that modes are well constrained.  Since 1795 pixels have one or more observations, it appears that only one signal mode, the mean, is unconstrained in the map.  For the case with the narrower beam separation, at low resolution again only the mean appears to be unconstrained, which supports the hypothesis that the bias we saw there was due to insufficient iterations at high $\lambda$ to debias the map before cooling.

We keep only the singular values of the well-constrained modes in our computation of the pseudoinverse.  The brute force solution in Fig.~\ref{fig:diffmess} shows smaller reconstruction errors than the messenger solution.   In Fig.~\ref{fig:diffmess} we also show the two eigenmodes of $(\mathbf{P}^\dag\mathbf{N}^{-1} \mathbf{P})$ with the largest eigenvalues and the two modes with the smallest eigenvalues (of the well-constrained modes).   The most constrained modes characterize the central region with the most crosslinking.  The least constrained modes are large scale gradients.  When we compute the solution with the pseudoinverse, we see that the remaining reconstruction error does indeed have a large scale gradient.

In the low resolution case, we also see that the messenger method with a $16 \times 10$ cooling schedule in Fig.~\ref{fig:diffmess} has not achieved reconstruction errors as small as the brute force solution.  On the other hand,  at 160 iterations it has a $\chi^2$ that is only a fraction $10^{-3}$ higher than the brute force solution (comparable to the conjugate gradient mapmaker's quality in the single beam case), so the large scale feature is not very significant.  Letting it continue to iterate at $\lambda = 1$ until 320 iterations lowers the $\chi^2$ steadily to a fraction $2 \times 10^{-4}$ above the brute force value.
We tested several fixed and adaptive cooling schedules.  The quickest we achieved a $\chi^2$ that is a fraction $10^{-3}$ above the brute force value is about 110 iterations, although that solution has larger large scale noise residuals than what we see in Fig.~\ref{fig:diffmess}.  A $\chi^2$ fraction $10^{-4}$ above the brute force value we can achieve after 670 iterations or so.  The slowest and most stringent case we tried reached a fraction $2 \times 10^{-5}$ above the brute force $\chi^2$ value after about 2700 iterations.



For many applications, a small number of messenger iterations may produce a map of sufficient quality.  It is also possible that a more clever approach to the cooling schedule could improve these results.

\section{Discussion of large surveys} \label{sec:discussion}

We discuss two points that are important in applying this method to the large detector arrays in current and future CMB experiments.  First, a practical algorithm should avoid storing the time-ordered data in memory, and second, it must treat the diversity of detector noise properties inherent in a multi-season ground-based CMB observing campaign.

\subsection{Projecting the data to map-sized objects}
To minimize its memory footprint, a mapmaking code should ideally avoid storing the full time-ordered data from iteration to iteration, and the computer should not read the data in more than once to avoid costly input/output operations.

In equation (\ref{eq:mapsized}) we showed that the messenger algorithm depends on the data only through the map-sized object $m_d(\lambda)$, which is far smaller than the full data.  If we establish a cooling schedule with levels $\{\lambda_j\}$ in advance, we can load the time-ordered data once from disk (in chunks), precompute the set of $\{m_d(\lambda_j)\}$ maps, and then flush the data from memory.  (In parallelization schemes where segments of the time-order data are distributed among computational nodes, each node needs only the parts of the $m_d(\lambda)$ maps that the data touch.)  This limitation to fixed $\lambda$-levels constrains the on-the-fly adaptability of our cooling schemes, but not the number of iterations per level.  We have shown already that simple cooling schemes can work well.

For real data sets, we do not want to compute $\chi^2(m_i)$ per iteration because it too requires the full data.  However, in our single beam test case we have shown that the behavior of $\chi^2$ per mode for individual detectors can be indicative of the convergence overall.  The $\chi^2$ per mode for representative detector timestreams thus can be a useful online diagnostic.

\subsection{Heterogeneous data}
Over the course of a multi-season campaign, differing observing conditions and loading will alter the minimum white noise level for detectors.  If $\mathbf{T}=\tau \mathbf{I}$, then $\tau$ must represent the global minimum noise for all timestreams going into the map.  This same argument applies to mixing timestreams from observatories with heterogeneous experimental designs, as in the current planning for the Simons Observatory and CMB-S4.  This single $\tau$ value will be too low for most detectors, and cause the convergence for them to go slower than it could have.  A better choice for $\mathbf{T}$ is piecewise proportional to the identity matrix, and uses appropriate $\tau$ values for each detector and each uncorrelated time-segment of the data:
\begin{eqnarray}
  \mathbf{T} &= {\rm diag}(
  &\tau_1,\dots,\tau_1, \\ \nonumber
  &&\tau_2,\dots,\tau_2, \\ \nonumber
  && \dots  \\ \nonumber
  &&\tau_{n_{\rm det}\times n_{\rm tod} },\dots,\tau_{n_{\rm det}\times n_{\rm tod} }).
\end{eqnarray}

In general, good choices for $\mathbf{T}$ can be best understood in terms of a hierarchical forward model for the time-ordered data:
\begin{itemize}
\item  Map $m$ is the sky signal.
\item  Messenger field $t$ is $\mathbf{P}m$ plus noise drawn from $\mathbf{T}$. 
\item  Data $d$ is $t$ plus noise drawn from $\bar \mathbf{N}$, so that the total amount of noise in $d$ is $\mathbf{N}$.
\end{itemize}
The more noise that we can put into $\mathbf{T}$, the better the method will work (while keeping $\bar \mathbf{N} = \mathbf{N} - \mathbf{T}$ positive definite). In the limit that all the noise is in $\mathbf{T}$, then $t = d$ and the method ``converges'' in one single step.  In practice we cannot do that because we cannot invert  $(\mathbf{P}^\dag \mathbf{T}^{-1} \mathbf{P})$ when $\mathbf{T}=\mathbf{N}$ for realistic cases, as it reverts to the standard mapmaking problem.

So our goal for $\mathbf{T}$ is that it should contain as much of the noise detail as possible, while still allowing for simple matrix inversions of $\bar \mathbf{N}$, $(\bar \mathbf{N}^{-1} +  (\lambda \mathbf{T})^{-1})$, $(\lambda^{-1} \bar \mathbf{N} +  \mathbf{T})$, and $(\mathbf{P}^\dag \mathbf{T}^{-1} \mathbf{P})$.  Uniform noise per detector timestream in $\mathbf{T}$ fulfills this goal.

\section{Conclusions}\label{sec:conclusions}
We have presented a mapmaking algorithm based on messenger fields.  The procedure is faster, and by several indications, produces higher quality maps than preconditioned conjugate gradient descent mapmakers, at least in our example.  It requires no preconditioner, but it does require a cooling schedule for rapid convergence.  We showed in our test cases that simple cooling schedules suffice.  A straightforward modification to the method allows treatment of differential data and deconvolution mapmaking.
 
We discussed ways in which the algorithm can scale up to handle real data sets.  We have successfully applied the messenger mapmaking method to polarization data from the Atacama Cosmology Telescope's ACTPol receiver \citep{2010SPIE.7741E..1SN},  adapting some of the existing mapmaking infrastructure to map one of the sky regions from \citet{2014JCAP...10..007N}.  We will discuss it more fully in future work, but our early efforts indicate that the map quality can be similar to conjugate gradient descent methods in a fraction of the iterations.  We have not worked hard yet to optimize the cooling schedule for the ACTPol data.  Like our simulation example here, the  preliminary map of the real data appears more fully converged on large scales than its conjugate gradient counterpart.  If these preliminary indications bear out, messenger field mapmaking may be a promising and powerful new addition to our CMB analysis toolbox.

\section*{Acknowledgments}
We thank Thibaut Louis, Arthur Kosowsky, Lyman Page, and Suzanne Staggs for useful comments and encouragement.
KMH acknowledges support from NASA grant NNX17AF87G.
The Flatiron Institute is supported by the Simons Foundation.

\bibliographystyle{hapj}
\bibliography{ref}

\appendix
\section{Derivation of messenger mapmaking} \label{sec:derivation}

The mapmaking equation minimizes 
\begin{equation}
  \chi^2(m,d) = (d - \mathbf{P} m)^\dag \mathbf{N}^{-1} (d - \mathbf{P} m),
\end{equation}
and maximizes the Gaussian probability $P(m|d) \propto \exp(-\chi^2(m,d)/2)$.  We introduce the augmented $\chi^2$ with a messenger field 
\begin{equation}
\chi_1^2(m,t,d) =  (t - \mathbf{P} m)^\dag \mathbf{T}^{-1} (t - \mathbf{P} m) + 
 (d - t)^\dag \bar \mathbf{N}^{-1} (d - t)
\end{equation}
(with $ \bar \mathbf{N} =  \mathbf{N} - \mathbf{T}$) that corresponds to the Gaussian probability $P_1(m,t|d) \propto \exp(-\chi_1^2(m,t,d)/2)$ and has the property that
\begin{equation}
  \int P_1 (m,t|d) dt = P(m|d)
\end{equation}
when $t$ is marginalized.  The field $t$ is called the \textit{messenger} field because it appears both with the map and the data in $\chi^2_1$, but the map and the data never appear together.  Thus $t$ communicates between the two.  The value of $m$ at the maximum is the same in both distributions $P$ and $P_1$.  By iteratively maximizing the marginal distributions $P_1(t|m,d)$ and $P_1(m|t,d)$, we can work our way up to the peak of $P_1(m,t|d)$, and find the $m$ that solves the mapmaking equation.

We can write down the marginal distributions by completing the square in $\chi_1^2$ for $t$ and $m$ respectively,
\begin{eqnarray}
  &\chi_1^2(t|m,d) &= (t - \mu_t)^\dag (\bar \mathbf{N}^{-1} + \mathbf{T}^{-1}) (t - \mu_t) + \dots \\ \nonumber
&\chi_1^2(m|t,d) &= (m - \mu_m)^\dag \mathbf{P}^\dag \mathbf{T}^{-1} \mathbf{P} (m - \mu_m) + \dots ,
\end{eqnarray}
and finding the maximum probability points (the means) of the marginal distributions at
\begin{eqnarray}
  \mu_t &=& \left( \bar \mathbf{N}^{-1} +  \mathbf{T}^{-1} \right)^{-1} \left(\bar \mathbf{N}^{-1} d + \mathbf{T}^{-1} \mathbf{P} m \right) \\ \nonumber
  \mu_m &=& (\mathbf{P}^\dag  \mathbf{T}^{-1} \mathbf{P})^{-1} \mathbf{P}^\dag \mathbf{T}^{-1} t.
\end{eqnarray}
Setting $t = \mu_t$ and $m = \mu_m$ leads to the iterative solution.  We are free to choose $\mathbf{T}$ so long as $\bar \mathbf{N}$ is positive definite.  Smart choices make the matrix inverses in the means of the marginal distributions easy to calculate.

If $\mathbf{T} = \tau \mathbf{I}$, then the matrix inverses are easy in the upper equation and the $\mathbf{T}$ cancels from the lower equation.  Alternatively, $\mathbf{T}$ can be made diagonal so that pieces of the time-order data each get their own optimized $\tau$ values, in which case $m$ in the next iteration is an weighted average of those pieces of the messenger field. 

\section{Mapmaking with a composite pointing matrix}


We can treat cases with a nontrivial pointing matrix by using a modified formalism where the full pointing matrix is a sum of simpler pointing matrices, each of which includes only a single pixel entry per time sample.  The case with two components has the form
\begin{equation}
  \mathbf{P} = \mathbf{P}_A + \mathbf{P}_B,
\end{equation}
but this could be generalized to an arbitrary number of components.  In the differential case  the $A$-component could have $+1$ entries and the $B$-component could have $-1$ entries.  The messenger field covariance will also be split, $\mathbf{T} = \mathbf{T}_A + \mathbf{T}_B$.  Matrices like $\mathbf{P}_A^\dag\mathbf{T}_A^{-1}\mathbf{P}_A$ can be made simple to invert.

The time-ordered data has the form
\begin{equation}
  d = \mathbf{P}_A m+ \mathbf{P}_B m + n.
\end{equation}
As before, we write down an augmented $\chi^2$, but this time with two messenger fields:
\begin{equation}
  \chi_1^2(m,t_A,t_B,d) = (t_A - \mathbf{P}_A m )^\dag \mathbf{T}_A^{-1} (t_A - \mathbf{P}_A m )
  + (t_B - \mathbf{P}_B m )^\dag \mathbf{T}_B^{-1} (t_B - \mathbf{P}_B m )
  +  (d - t_A - t_B)^\dag \bar \mathbf{N}^{-1} (d - t_A - t_B).
\end{equation}
Each messenger field has an associated covariance matrix $\mathbf{T}_A,\mathbf{T}_B$.  With few restrictions, we are able to choose these matrices to make subsequent operations simpler.  Again we have $ \bar \mathbf{N} =  \mathbf{N} - \mathbf{T} =  \mathbf{N} - \mathbf{T}_A - \mathbf{T}_B$.

As before we complete the square to find the means of the marginal distributions,
\begin{eqnarray}
  \chi_1^2(t_A| \dots) &=&  (t_A - \mu_{t_A})^\dag (\bar \mathbf{N}^{-1} + \mathbf{T}_A^{-1}) (t_A - \mu_{t_A}) + \dots \\ \nonumber
  \chi_1^2(t_B| \dots) &=&  (t_B - \mu_{t_B})^\dag (\bar \mathbf{N}^{-1} + \mathbf{T}_B^{-1}) (t_B - \mu_{t_B}) + \dots \\ \nonumber
  \chi_1^2(m| \dots) &=& (m - \mu_m)^\dag ( \mathbf{P}_A^\dag \mathbf{T}_A^{-1}\mathbf{P}_A + \mathbf{P}_B^\dag \mathbf{T}_B^{-1}\mathbf{P}_B ) (m - \mu_m) + \dots ,
\end{eqnarray}
and find
\begin{eqnarray} \label{eqn:composite_pointing_means} 
  \mu_{t_A} &=& \left( \bar \mathbf{N}^{-1} +  \mathbf{T}_A^{-1} \right)^{-1} \left(\bar \mathbf{N}^{-1} (d-t_B) + \mathbf{T}_A^{-1} \mathbf{P}_A m \right) \\ \nonumber
  \mu_{t_B} &=& \left( \bar \mathbf{N}^{-1} +  \mathbf{T}_B^{-1} \right)^{-1} \left(\bar \mathbf{N}^{-1} (d-t_A) + \mathbf{T}_B^{-1} \mathbf{P}_B m \right) \\ \nonumber
  \mu_m &=&  ( \mathbf{P}_A^\dag \mathbf{T}_A^{-1}\mathbf{P}_A + \mathbf{P}_B^\dag \mathbf{T}_B^{-1}\mathbf{P}_B )^{-1} ( \mathbf{P}_A^\dag \mathbf{T}_A^{-1}t_A + \mathbf{P}_B^\dag \mathbf{T}_B^{-1}t_B ).
\end{eqnarray} 
The iterative solution follows from $t_A = \mu_{t_A}$, $t_B = \mu_{t_B}$, and $m = \mu_m$.  These equations require inversions only of trivial matrices and are sufficient to find the mapmaking solution.  The term $( \mathbf{P}_A^\dag \mathbf{T}_A^{-1}\mathbf{P}_A + \mathbf{P}_B^\dag \mathbf{T}_B^{-1}\mathbf{P}_B )$ is a weighted sum of weight maps (summed over pointing component) when we choose $\mathbf{T}_A,\mathbf{T}_B$ that are piecewise proportional to the identity matrix.  The generalization to more pointing components is straightforward.

Despite the appearance of $t_A$ and $t_B$, we can write the solution in terms of a single messenger field $t = t_A + t_B$.  Then $t_B = t - t_A$ and vice versa.  In that case
\begin{equation}
  t_A = \left( \bar \mathbf{N}^{-1} +  \mathbf{T}_A^{-1} \right)^{-1}  \left(\bar \mathbf{N}^{-1} (d-t) + \bar \mathbf{N}^{-1} t_A + \mathbf{T}_A^{-1} \mathbf{P}_A m \right),
\end{equation}
and similar for $t_B$.  Putting all the $t_A$ terms on the left hand side, and multiplying both sides by $\mathbf{T}_A ( \bar \mathbf{N}^{-1} +  \mathbf{T}_A^{-1})$, yields
\begin{equation}
  t_A =  \mathbf{T}_A \bar \mathbf{N}^{-1} (d-t) +  \mathbf{P}_A m, \label{eqn:tAproj}
\end{equation}
and similar for $t_B$.  If we add the two messenger fields together we find that
\begin{equation}
  t = ( \mathbf{T}_A +  \mathbf{T}_B)  \mathbf{N}^{-1} d - ( \mathbf{T}_A +  \mathbf{T}_B)  \mathbf{N}^{-1} t + (\mathbf{P}_A + \mathbf{P}_B) m
\end{equation} which can be rearranged to give
\begin{equation}
  t =  \left( \bar \mathbf{N}^{-1} +  \mathbf{T}^{-1} \right)^{-1} \left(\bar \mathbf{N}^{-1} d + \mathbf{T}^{-1} \mathbf{P} m \right),
\end{equation}
the same as the case with a simple pointing matrix.  Thus we conclude that there is only one real messenger field, namely $t$, and $t_A$ and $t_B$ are simply projections of it (using equation \ref{eqn:tAproj}).

We can plug those projections into the the final term in the expression for the map in equation (\ref{eqn:composite_pointing_means}), and work to express the right-hand side in terms of the data and map only:
\begin{eqnarray}
  \mathbf{P}_A^\dag \mathbf{T}_A^{-1}t_A + \mathbf{P}_B^\dag \mathbf{T}_B^{-1}t_B =& & 
  \mathbf{P}_A^\dag \bar \mathbf{N}^{-1} (d-t) +  \mathbf{P}_A^\dag \mathbf{T}_A^{-1} \mathbf{P}_A m \\ \nonumber
  &&+  \mathbf{P}_B^\dag \bar \mathbf{N}^{-1} (d-t) +  \mathbf{P}_B^\dag \mathbf{T}_B^{-1} \mathbf{P}_B m \\ \nonumber
  =&& \mathbf{P}^\dag \bar \mathbf{N}^{-1} (d-t) + ( \mathbf{P}_A^\dag \mathbf{T}_A^{-1}\mathbf{P}_A + \mathbf{P}_B^\dag \mathbf{T}_B^{-1}\mathbf{P}_B ) m \\ \nonumber
  =&& \mathbf{P}^\dag \bar \mathbf{N}^{-1} d \\ \nonumber
  &&-  \mathbf{P}^\dag \bar \mathbf{N}^{-1}  \left( \bar \mathbf{N}^{-1} +  \mathbf{T}^{-1} \right)^{-1} \left(\bar \mathbf{N}^{-1} d + \mathbf{T}^{-1} \mathbf{P} m \right) \\ \nonumber
  &&+ ( \mathbf{P}_A^\dag \mathbf{T}_A^{-1}\mathbf{P}_A + \mathbf{P}_B^\dag \mathbf{T}_B^{-1}\mathbf{P}_B ) m \\ \nonumber
  =&& \mathbf{P}^\dag (\bar \mathbf{N}^{-1} - \bar \mathbf{N}^{-1}  \left( \bar \mathbf{N}^{-1} +  \mathbf{T}^{-1} \right)^{-1} \bar \mathbf{N}^{-1} ) d \\ \nonumber
  &&+ \left(
  \mathbf{P}_A^\dag \mathbf{T}_A^{-1}\mathbf{P}_A + \mathbf{P}_B^\dag \mathbf{T}_B^{-1}\mathbf{P}_B )
  -  \mathbf{P}^\dag \bar \mathbf{N}^{-1}  \left( \bar \mathbf{N}^{-1} +  \mathbf{T}^{-1} \right)^{-1} \mathbf{T}^{-1} \mathbf{P}
  \right) m \\ \nonumber
  =&&  \mathbf{P}^\dag ( \bar \mathbf{N} +  \mathbf{T} )^{-1} d +  \left(
  (\mathbf{P}_A^\dag \mathbf{T}_A^{-1}\mathbf{P}_A + \mathbf{P}_B^\dag \mathbf{T}_B^{-1}\mathbf{P}_B )
  -  \mathbf{P}^\dag ( \bar \mathbf{N} +  \mathbf{T} )^{-1}  \mathbf{P}
  \right) m.
\end{eqnarray}
We used the Woodbury formula in the first term of the last step.

Multiplying by $(\mathbf{P}_A^\dag \mathbf{T}_A^{-1}\mathbf{P}_A + \mathbf{P}_B^\dag \mathbf{T}_B^{-1}\mathbf{P}_B )$, we arrive at a compact expression for updating the map iterations:
\begin{eqnarray}
  m_{i+1} &=&   (\mathbf{P}_A^\dag \mathbf{T}_A^{-1}\mathbf{P}_A + \mathbf{P}_B^\dag \mathbf{T}_B^{-1}\mathbf{P}_B )^{-1}  \mathbf{P}^\dag ( \bar \mathbf{N} +  \mathbf{T} )^{-1} d \\ \nonumber
  && + (\mathbf{I} -  (\mathbf{P}_A^\dag \mathbf{T}_A^{-1}\mathbf{P}_A + \mathbf{P}_B^\dag \mathbf{T}_B^{-1}\mathbf{P}_B )^{-1}  \mathbf{P}^\dag ( \bar \mathbf{N} +  \mathbf{T} )^{-1} \mathbf{P} ) m_i \\ \nonumber
  &=& m_d + \mathbf{F} m_i.
\end{eqnarray}
This expresses the mapmaking solution in a way that requires only map-sized objects be stored in memory.  The expressions 
\begin{eqnarray}
   m_d &=& (\mathbf{P}_A^\dag \mathbf{T}_A^{-1}\mathbf{P}_A + \mathbf{P}_B^\dag \mathbf{T}_B^{-1}\mathbf{P}_B )^{-1}  \mathbf{P}^\dag ( \bar \mathbf{N} +  \mathbf{T} )^{-1} d \\ \nonumber
  \mathbf{F} &=& \mathbf{I} -  (\mathbf{P}_A^\dag \mathbf{T}_A^{-1}\mathbf{P}_A + \mathbf{P}_B^\dag \mathbf{T}_B^{-1}\mathbf{P}_B )^{-1}  \mathbf{P}^\dag ( \bar \mathbf{N} +  \mathbf{T} )^{-1} \mathbf{P}
\end{eqnarray}
recall similar ones in equations (\ref{eqn:md}) and (\ref{eqn:F}) for the single beam case.  Including the $\lambda$ parameters, we have
\begin{eqnarray}
   m_d(\lambda) &=& (\mathbf{P}_A^\dag \mathbf{T}_A^{-1}\mathbf{P}_A + \mathbf{P}_B^\dag \mathbf{T}_B^{-1}\mathbf{P}_B )^{-1}  \mathbf{P}^\dag (\lambda^{-1} \bar \mathbf{N} +  \mathbf{T} )^{-1} d \\ \nonumber
  \mathbf{F}(\lambda) &=& \mathbf{I} -  (\mathbf{P}_A^\dag \mathbf{T}_A^{-1}\mathbf{P}_A + \mathbf{P}_B^\dag \mathbf{T}_B^{-1}\mathbf{P}_B )^{-1}  \mathbf{P}^\dag ( \lambda^{-1}\bar \mathbf{N} +  \mathbf{T} )^{-1} \mathbf{P}
\end{eqnarray}

Another convenient expression of the solution is
\begin{equation}
   m_{i+1} =  m_{i} + (\mathbf{P}_A^\dag \mathbf{T}_A^{-1}\mathbf{P}_A + \mathbf{P}_B^\dag \mathbf{T}_B^{-1}\mathbf{P}_B )^{-1}  \mathbf{P}^\dag ( \lambda^{-1}\bar \mathbf{N} +  \mathbf{T} )^{-1} (d - \mathbf{P}m_i),
\end{equation}
which recalls equation (\ref{eqn:mcompact}).

Finally, we can check that we recover the simple, single pencil beam case by setting $\mathbf{P}_B = 0$.  Then $\mathbf{P} = \mathbf{P}_A$ and  $ \bar \mathbf{N}_A =  \bar \mathbf{N} + \mathbf{T}_B = \mathbf{N} - \mathbf{T}_A$.  So we have
\begin{eqnarray}
  m_d &=& (\mathbf{P}^\dag \mathbf{T}_A^{-1}\mathbf{P} )^{-1}  \mathbf{P}^\dag (\lambda^{-1} \bar \mathbf{N}_A +  \mathbf{T}_A )^{-1} d \\ \nonumber
   \mathbf{F} &=&  \mathbf{I} -  (\mathbf{P}^\dag \mathbf{T}_A^{-1}\mathbf{P} )^{-1}  \mathbf{P}^\dag (\lambda^{-1}\bar \mathbf{N}_A +  \mathbf{T}_A )^{-1} \mathbf{P}
\end{eqnarray}
and recover one of the forms for the single pencil beam case from before  (equations \ref{eqn:md} and \ref{eqn:F}).

\section{Mapmaking with composite timestream noise} \label{sec:2mess}
The timestream noise covariance could be complicated and composed of several pieces, like $\mathbf{N} = \sum_i \mathbf{N}_i$, so that it by itself is difficult to invert in any easily accessible basis.  If the sub-components are invertible in separate and convenient bases, \citet{2017arXiv170400865H} showed that the messenger field method can be extended to handle such cases.  That work examined Wiener filters, but the same idea applies for mapmaking.  For the case $\mathbf{N} = \mathbf{N}_0 + \mathbf{N}_1$ we need additional messenger fields $t_0,t_1,d_0$ and two matrices $\mathbf{T}_0,\mathbf{T}_1$.  (More complicated noise will require even more additional fields.)
Iterating the following equations will yield the mapmaking solution: 
\begin{eqnarray}
   t_1 &=&(\bar  \mathbf{N}_1^{-1} +  \mathbf{T}_1^{-1})^{-1} \left( \bar  \mathbf{N}_1^{-1} d + \mathbf{ T}_1^{-1} d_0 \right)  \label{eqn:2mess_t1}  \\ \nonumber
   d_0 &=& (\bar  \mathbf{N}_0^{-1} +  \mathbf{T}_1^{-1})^{-1} \left( \mathbf{ \bar N}_0^{-1} t_0 + \mathbf{ T}_1^{-1} t_1 \right) \label{eqn:2mess_d0} \\ \nonumber
   t_0 &=& (\bar  \mathbf{N}_0^{-1} +  \mathbf{T}_0^{-1})^{-1} \left( \mathbf{ \bar N}_0^{-1} d_0 + \mathbf{ T}_0^{-1} \mathbf{P} m \right) \label{eqn:2mess_t0} \\  \nonumber
   m &=& (\mathbf{P}^\dag\mathbf{T}_0^{-1} \mathbf{P})^{-1} \mathbf{P}^\dag \mathbf{T}_0^{-1} t_0,
   \label{eqn:2mess_m}
\end{eqnarray}
where $\bar \mathbf{N}_0 = \mathbf{N}_0 - \mathbf{T}_0$ and similarly for $\bar \mathbf{N}_1$.

Gap filling, for example, can be naturally incorporated into this scheme.  We can let $\mathbf{N}_0$ be the timestream noise (diagonal in the Fourier domain) and let $\mathbf{N}_1$ have infinite variance during gaps in the timestream (diagonal in the time domain) but minimal other noise.  This is similar to the treatment of spatial masks in \citet{2013A&A...549A.111E}, \citet{2017arXiv170400865H}, and many other works.  Under this assumption, $d_0$ will be an optimized estimate of the gap-filled timestream and there are no ambiguities about how to fill gaps.   However, this method may not be practical because of the need to carry around additional messenger fields.  Each of these is large for real data sets, the same size as the time-ordered data.
The additional fields will also make the system slower to converge.

\end{document}